\documentclass[titlepage, 12pt]{article}

\usepackage[round]{natbib} 

\usepackage{amsmath, amsbsy, epsfig, epsf, psfrag, booktabs, graphicx, amssymb}

\usepackage{lscape}

\usepackage[margin=1in]{geometry}

\usepackage{color}
\usepackage{authblk}

\usepackage[pagewise]{lineno}
    \newcommand*\patchAmsMathEnvironmentForLineno[1]{%
      \expandafter\let\csname old#1\expandafter\endcsname\csname #1\endcsname
      \expandafter\let\csname oldend#1\expandafter\endcsname\csname end#1\endcsname
      \renewenvironment{#1}%
         {\linenomath\csname old#1\endcsname}%
         {\csname oldend#1\endcsname\endlinenomath}}%
    \newcommand*\patchBothAmsMathEnvironmentsForLineno[1]{%
      \patchAmsMathEnvironmentForLineno{#1}%
      \patchAmsMathEnvironmentForLineno{#1*}}%
    \AtBeginDocument{%
    \patchBothAmsMathEnvironmentsForLineno{equation}%
    \patchBothAmsMathEnvironmentsForLineno{align}%
    \patchBothAmsMathEnvironmentsForLineno{flalign}%
    \patchBothAmsMathEnvironmentsForLineno{alignat}%
    \patchBothAmsMathEnvironmentsForLineno{gather}%
    \patchBothAmsMathEnvironmentsForLineno{multline}%
    }

\begin{document}

\title{Modelling death rates due to COVID-19: A Bayesian approach}

\renewcommand\Affilfont{\normalfont\small}
\author{Cristian Bayes}
\author{Giancarlo Sal y Rosas\footnote{Corresponding author: vsalyrosas@pucp.edu.pe}}
\author{Luis Valdivieso}
\affil{Departamento de Ciencias, Pontificia Universidad Cat\'olica del Per{\'u}}

\maketitle

\begin{abstract}
{\bf Objective}: To estimate the number of deaths in Peru due to COVID-19. 
Design: With a priori information obtained from the daily number of deaths due to CODIV-19 in China and data from the Peruvian authorities, we constructed a predictive Bayesian non-linear model for the number of deaths in Peru. \\
{\bf Exposure}: COVID-19. \\
{\bf Outcome}: Number of deaths. \\
{\bf Results}: Assuming an intervention level similar to the one implemented in China, the total number of deaths in Peru is expected to be 612 (95\%CI: 604.3 - 833.7) persons. Sixty four days after the first reported death, the 99\% of expected deaths will be observed. The inflexion point in the number of deaths is estimated to be around day 26 (95\%CI: 25.1 - 26.8) after the first reported death. \\
{\bf Conclusion}: These estimates can help authorities to monitor the epidemic and implement strategies in order to manage the COVID-19 pandemic. 
\end{abstract}

%
%

\section{Introduction}

There is a trend to forecast COVID-19 using mathematical models for the probability of moving between states from susceptible to infected, and then to a recovered state
or death (SIR models). This approach is however very sensitive to starting assumptions and tend to overestimated the virus reproductive rate. One key point that these models miss is the  individual behavioral responses and government-mandated policies that  can dramatically influence
the course of the epidemic. In Wuhan, for instance, strict social distancing was instituted on January 23rd, 2020, and by March 15th new infections were close to zero.  Taking into account this observation, 
\citet{covid2020forecasting} have proposed  a statistical approach to model a empirical cumulative population death rate. However, modelling observed cumulative death numbers has the inherent problem of yielding a  highly correlated data which, if it is not taken into account, could cause misleading  inference results.  

Instead of considering a mathematical model such as SIR, that rely heavily on parameters assumptions, we will directly work as in \citet{covid2020forecasting} or \citep[]{zhou2020forecasting} with  empirical data. To this end,  we propose to model the daily number of deaths using a Poisson distribution with a rate parameter that is proportional to a  Skew Normal density. Using a Bayesian approach and a prior epidemic China COVID-19 history, we forecast the total number of deaths in Peru for the next seventy days. 
%
%
 
\section{Methods}

\subsection{Model}

Let  $Y_{t_1}, Y_{t_2}, \ldots, Y_{t_n}$ be the number of COVID-19 deaths at times $t_1, t_2, \ldots, t_n$, where time is measured from the first reported death due to COVID-19, and let us suppose that the death rate will hit a platoon due to the government intervention. We propose then the model

\begin{eqnarray}
Y(t_i) &\sim& Poisson(\lambda(t_i)), \ i =1,2,\ldots,n
\label{eq1} 
\end{eqnarray}
with  death rate
\begin{eqnarray}
\lambda(t_i) &=& p g(t_i \mid  \alpha, \beta, \eta)K,
\label{eq2}
\end{eqnarray}

\noindent where $g(t_i \mid  \alpha, \beta, \eta)$ denotes the density function of a Skew Normal distribution with location, scale, and shape parameters,  $\alpha$, $\beta$, and $\eta$, respectively; $p$ is a maximum asymptotic level parameter  and $K$  is the population size.  The choice of the  Skew Normal distribution  is motivated by its  flexibility on the tails and their asymmetry, which  can force, as one could  expect, a rapidly increase rate at the first stages of the pandemic and a slower decrease of this rate at the last stages. A  negative binomial distribution can also be considered for the deaths numbers, but we found empirically a better fit with the Poisson model.

Our model differs of the approach taken by  \citet{covid2020forecasting}, who directly models the cumulative death rate $\Lambda(t) = \int_0^t \lambda(s)ds$ with a term proportional to a symmetric cumulative normal distribution. This difference is not only found in the formulation, but also in the estimation procedure. While these authors incorporated first the Chinese data--more concretely the time from
when the initial death rate exceeds 1e-15 to the implementation of social distancing-- into their model throughout a location-specific inflection point parameter or a maximum death rate and then used a sort of credibility model between short-range and long-rate variants, we followed a Bayesian approach that incorporates the Chinese data as a prior distribution. The posterior predictive distribution, which is the goal of this model, is then a dynamic object that can be updated with new data   about the number of deaths or any other reliable information  that may be incorporated as covariates in the model.

Apart from the total number of deaths, three main quantities of interest  can be easily derived from our model. First,  a time to threshold death rate, which  provides  the time after which only 0.01 $\%$ of deaths will be observed on the population. This will be defined as the 0.99 quantile of the $g$ distribution. Another characteristic of interest is the inflection point, defined as the time at which the death rate reaches its maximum level.

\subsection{Priors}

\begin{figure}[!htb]
\begin{center}	\includegraphics[width=15.0cm,height=10.0cm]{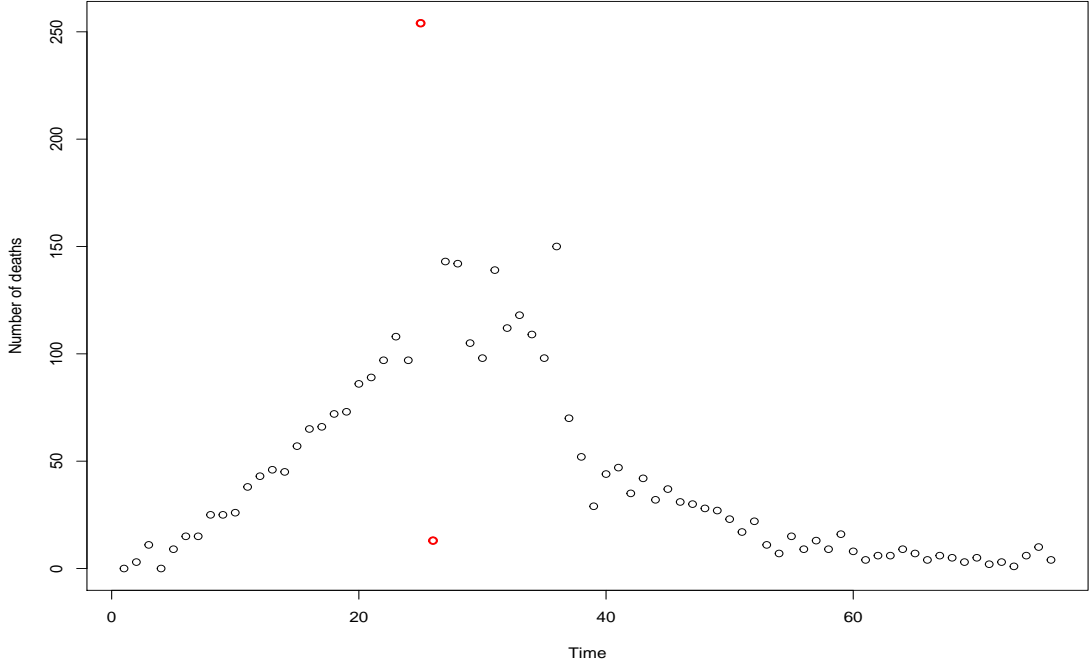}
\caption{Empirical distribution of number of deaths reported in China. Points in red represents the information on February 13 and 14, respectively.}\label{fig:lfig0}
\end{center}
\end{figure}

Since China was the first country to have experienced  a drastic drop in infections and deaths, we are proposing to incorporate this data, into our Peruvian death rate predictions, through a prior distribution. Figure \ref{fig:lfig0} shows the empirical distribution of number of reported deaths in China.  One can notice that the reported numbers on February 13 and 14 (red points) were 254 and 13 deaths, respectively. There is some controversy\footnote{https://www.bbc.com/news/world-51495484}  about the reported numbers from China on these days, the reason why we are considering  the average number for these days.

\begin{figure}[!htb]
\begin{center}	\includegraphics[width=15.0cm,height=10.0cm]{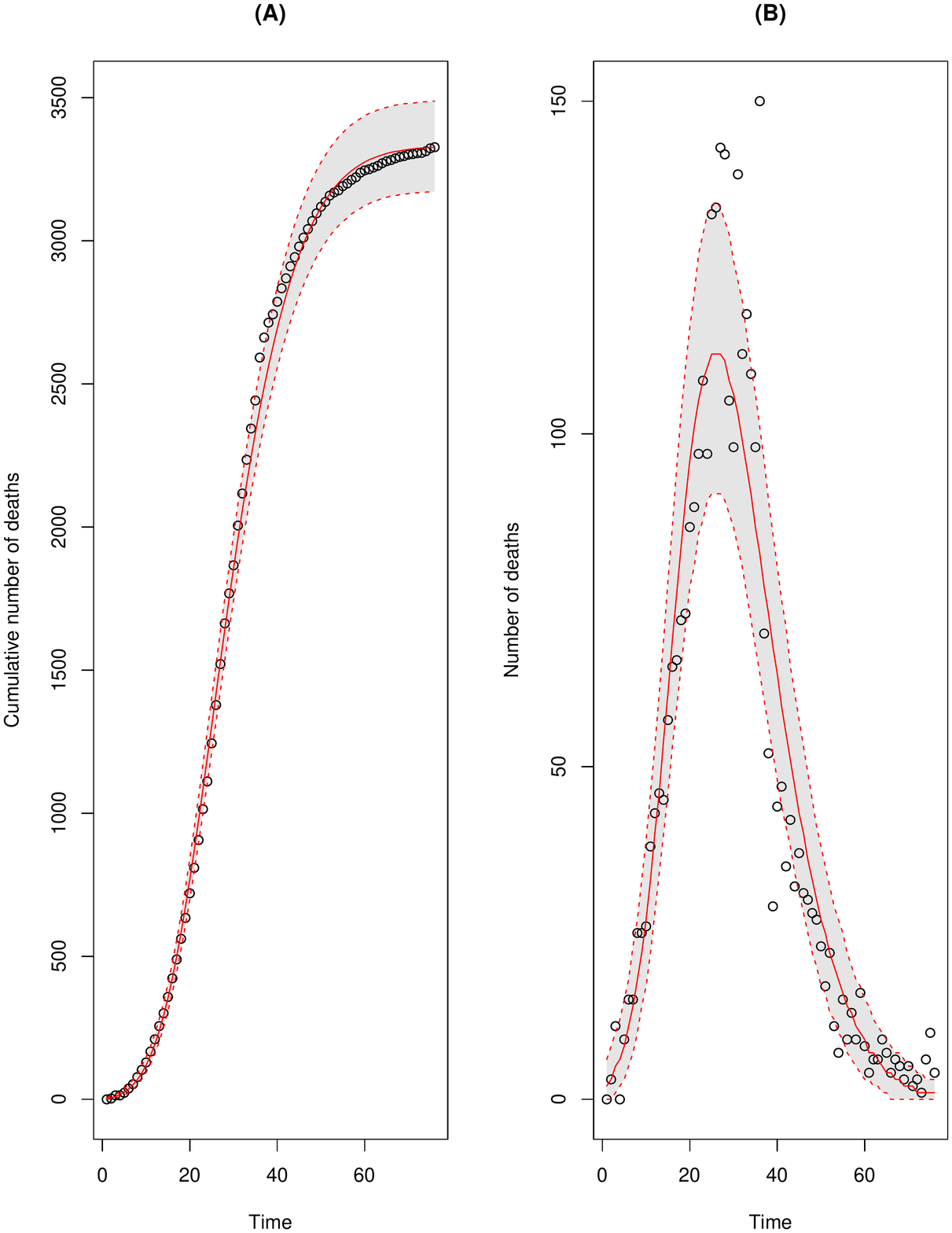}
\caption{Figure A: Estimated cumulative mortality rate for China and its associated 95\% predictive interval where points represent the empirical data. Figure B: Estimated number of infected for China and its associated 95\% predictive interval where points represent the empirical data.}\label{fig:lfig1}
\end{center}
\end{figure}

Figure \ref{fig:lfig1} shows the observed and predicted death rates (A) and the daily number of deaths (B) in China. The predicted rates and their associated 95\% prediction credible intervals were obtained, under a Bayesian approach, by considering a non-informative prior and the Chinese official death reports. Table \ref{tab:tab1} summarizes the information given in Figure \ref{fig:lfig1}. This information will be considered as a prior for modelling the number of deaths in Per\'u with the exception of the $p$  parameter, where a weakly informative prior, $N(0,10^2)$, will be considered for $\log{(p)}$. 

\begin{table}[ht]
\caption{China: Mean, standard deviation (SD) and quantiles 0.25,0.5 and 0.975 of the posterior distribution}
\begin{center}
\begin{tabular}{lccccc}
  \hline
 & Mean & SD & 2.5\% & 50\% & 97.5\% \\
 \hline 
 $\beta$ & 16.52 & 0.37 & 15.81 & 16.52 & 17.24 \\ 
  $\log{(\alpha)}$ & 2.91 & 0.02 & 2.87 & 2.91 & 2.95 \\ 
  $\eta$ & 2.34 & 0.14 & 2.07 & 2.34 & 2.64 \\ 
  $\log{(p)}$ & 0.87 & 0.02 & 0.84 & 0.87 & 0.91 \\ 
   \hline
\end{tabular}
\end{center}
\label{tab:tab1}
\end{table}

Initial values for the estimation process will be sampled from the prior distribution that was constructed with the Chinese data.

%
%

\subsection{Data}

The Chinese data to be used in this work was obtained from  the European Centre for Disease Prevention and Control, institution  that  daily publishes statistics on the COVID-19 pandemic. The daily death  reports in Per\'u, on the other hand,  were obtained from the local authorities.

%
%

\subsection{Inference}

Taking into account the observed likelihood function, easily derived from (\ref{eq1}) and (\ref{eq2}), the posterior distribution of the vector parameter $(p,\alpha,\beta,\eta)$ has a complex expression. To obtain samples from it, we will make use of Hamiltonian MCMC methods as the ones implemented in the R package RStan \citep{rstan}. This package implements an adaptive Hamiltonian Monte Carlo algorithm (also known as HMC) using a No-U-Turn sampler (NUTS) for the stepsize parameter in order to generate efficient transitions to the posterior distribution \citep*{carpenter2017stan}.

%
%

\section{Results: Forecasting for Per\'u}

We now briefly describe our predictions for the spread of COVID-19 in Per\'u. The time here will be understood to be measured in days after the first reported death in the country.

\subsection{Number of deaths}

\begin{figure}[!htb]
	\begin{center}	\includegraphics[width=15.0cm,height=10.0cm]{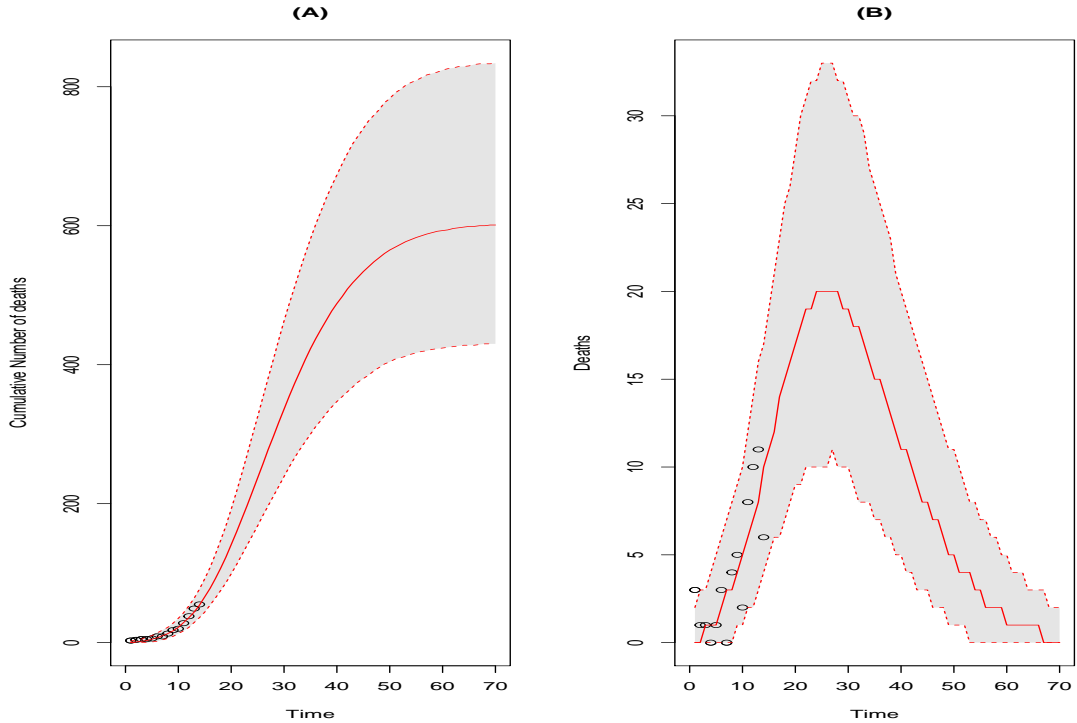}
		\caption{A: Estimated cumulative number of deaths for Per\'u and its associated 95\% predictive interval. B: Estimated number of deaths for Per\'u and its associated 95\% predictive  interval.}\label{fig:lfig2}
	\end{center}
\end{figure}

Figure \ref{fig:lfig2}.A shows the estimated cumulative number of deaths for Per\'u and its associated 95\% predictive credibility interval. In particular, the expected total number of deaths is 141.8 (95\%CI: 99 - 192) and 339.7 (95\% CI: 239 - 462) after 20 and 30 days, respectively.

In addition, Figure \ref{fig:lfig2}.B shows the expected number of deaths per day and its associated 95\% predictive credibility interval. In particular, the expected number of deaths is 17.5 (95\%CI: 9 - 28) and 19.3 (95\% CI: 10 - 31) at 20 and 30 days, respectively. Overall, the total number of deaths is expected to be 611.6 (95\%CI: 604.3 - 833.7) persons.

%
%

\subsection{Time to threshold death rate}

The model estimates that the number of days, since the first death, to achieve the threshold death rate  will be 63.8 (95\%CI: 52.9 - 65.84) days. At this time, the 99\% of expected deaths will have been observed.

\subsection{Inflection point}

The estimated inflection point is 25.9 (95\%CI: 25.1 - 26.8), which means that we expect to spend approximately 26 days before the COVID-19 death rate starts to decline after the first reported death case. 

%
%

%
%

\section{Sensitivity analysis}

Three different scenarios were considered, each keeping the weakly informative prior log-normal distribution for $p$. In all cases,  the posterior mean for the fitted Chinese model  was taken as the mean prior for the modelling of data from Per\'u. The first scenario (I) takes also the posterior Chinese variances as  the prior Peruvian variances. The second scenario (II) induces flexibility in the prior distribution by increasing their corresponding  standard deviations by a factor of five. Finally, the third scenario (III)  increases the standard deviations by a factor of ten. 

\begin{figure}[!htb]
	\begin{center}		\includegraphics[width=15.0cm,height=10.0cm]{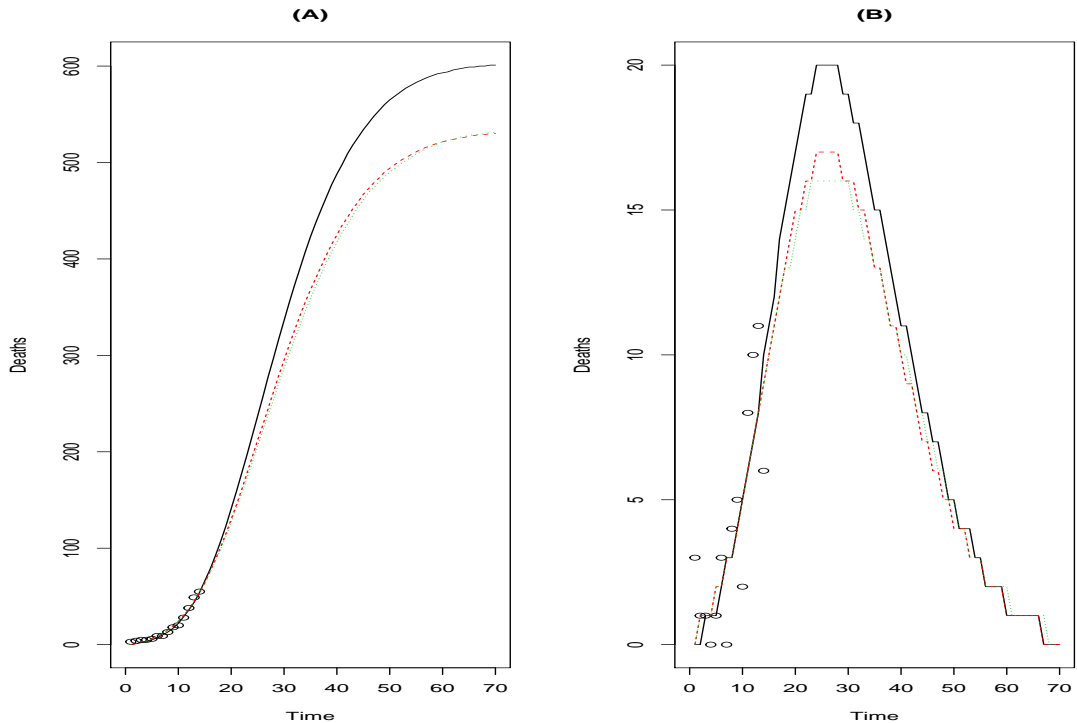}
		\caption{Predictive distribution under three scenarios. Scenario I: Prior distribution using Chinese information and $p$ free (black). Scenario II: Prior standard deviations of Scenario A were multiply by 5 (red). Scenario III: Prior standard deviations of Scenario A were multiply by 10 (green).}\label{fig:lfig4}
	\end{center}
\end{figure}

Table \ref{tab:tab2} and Figure \ref{fig:lfig4} shows that the point estimates are not heavily affected, but the precision pays the price of not having yet enough data from Per\'u. 

\begin{table}[ht]
\caption{Mean (95\% credible interval) of the posterior distribution under three scenarios. Scenario I: Prior distribution using the Chinese information and $p$ free, Scenario II: Same as Scenario I, but multiplying by 5 the standard deviations, and Scenario III: Same as Scenario I, but multiplying by 10 the standard deviations.}
\begin{center}
\begin{tabular}{cccc}
Scenario & Time to threshold & Inflection point & Total number of deaths \\
\hline
I        & 63.8 (61.9 - 65.8) & 26.0 (25.1 - 26.8) 
& 611.6 (437.2 - 833.7)\\
II & 64.7 (55.4 - 74.9) & 26.2 (22.0 - 30.3)
& 568.7 (299.4 - 1037.5) \\
III & 66.6 (48.9 - 88.5) & 26.6 (19.0 - 34.9) 
& 629.6 (225.8 - 1588.4 )\\
\hline
\end{tabular}
\end{center}
\label{tab:tab2}
\end{table}

%
%

\section{Discussion}

Considering the information on daily number of deaths from China and from Per\'u, our estimation of total number of deaths will be 611.6 (95\% CI: 437.2 - 833.7) and 99\% of those deaths will occur 63.8 (95\% CI: 52.9 - 65.8) days after the first death reported case.

The present study has some limitations. Although Per\'u has not followed all the measures taken  in China, we assumed that Peruvian interventions will have similar effects as the ones observed in that country. However, we expect our model is flexible enough to be driven by the Peruvian data. Furthermore, the model will increase its precision as more data from Per\'u becomes available.

Several extensions are possible for the model. For instance, covariates as daily social mobility indicators or country age distributions can be included in the model, in particular on the $p$ parameter that measures the total number of deaths.

We expect our model can be useful to guide some policies that need to be taken by the Peruvian government in order to overcome the COVID-19 pandemic. For example, the proposed model can be useful to measure the impact of the COVID-19 pandemic on the Peruvian health system.

\section{Acknowledgement}

We would like to thanks Rodrigo Carrillo Larco for providing us the detailed information of Per\'u based on reports by the Peruvian Ministry of Health and Dr. Daniel Manrique-Vallier for earlier discussions and helpful suggestions on the topic. 


\bibliography{bibliografia}
\bibliographystyle{dcu}

\end{document}